# Merging toroidal dipole bound states in the continuum without up-down symmetry in Lieb lattice metasurfaces

Guodong Zhu[1,3], Sen Yang[1,2] and Justus C. Ndukaife[1,3*]
[1]Vanderbilt Institute of Nanoscale Science and Engineering, Vanderbilt University, Nashville, TN, 37235, USA
[2]Interdisciplinary Materials Science, Vanderbilt University, Nashville, TN, 37235, USA
[3]Department of Electrical Engineering and Computer Science, Vanderbilt University, Nashville, TN, 37235, USA
Corresponding author: Justus C. Ndukaife justus.ndukaife@vanderbilt.edu

**Abstract**
The significance of bound states in the continuum (BICs) lies in their potential for theoretically infinite quality factors. However, their actual quality factors are limited by imperfections in fabrication, which lead to coupling with the radiation continuum. In this study, we present a novel approach to address this issue by introducing a merging BIC regime based on a Lieb lattice. By utilizing this approach, we effectively suppress the out-of-plane scattering loss, thereby enhancing the robustness of the structure against fabrication artifacts. Notably, unlike previous merging systems, our design does not rely on the up-down symmetry of metasurfaces. This characteristic grants more flexibility in applications that involve substrates and superstrates with different optical properties, such as microfluidic devices. Furthermore, we incorporate a lateral band gap mirror into the design to encapsulate the BIC structure. This mirror serves to suppress the in-plane radiation resulting from finite-size effects, leading to a remarkable ten-fold improvement in the quality factor. Consequently, our merged BIC metasurface, enclosed by the Lieb lattice photonic crystal mirror, achieves an exceptionally high-quality factor of $10^5$ while maintaining a small footprint of 26.6×26.6 um. Our findings establish an appealing platform that capitalizes on the topological nature of BICs within compact structures. This platform holds great promise for various applications, including optical trapping, optofluidics, and high-sensitivity biodetection, opening up new possibilities in these fields.

## 1. Introduction

Photonic bound states in the continuum (BICs) can confine light with a theoretically infinite lifetime, leading to a strong light-matter interaction. BICs were first proposed mathematically in quantum mechanics and then subsequently observed in optics and classical waves [1]. They have been identified as the topological defects of polarization vectors in reciprocal space [2]. To date, a variety of mechanisms are adopted to construct BICs including exploring structure parameter space, topological charge evolution as well as parity-time symmetry [3]. Generally, BICs are classified into symmetry-protected BICs and accidental BICs with the former obtained by structure symmetry and the latter obtained through parameter tuning [1]-[8]. Compared to Mie resonant dielectric nanoantenna systems, plasmonic nanostructures or photonic crystal cavities [9], the utilization of BICs in all-dielectric systems has emerged as a promising alternative in nanophotonics due to their ability to offer significantly high field enhancement and sharp resonance while avoiding the material absorption and thermal effects intrinsic to metal plasmonic systems. Applications based on BICs such as surface-enhanced Raman spectroscopy, fluorescence enhancement, biosensing, lasing, ultra-fast switching, hyperspectral imaging, etc. have been reported [10]-[17].

Nonetheless, practical BIC systems face challenges due to fabrication imperfections, disorders, and unavoidable material absorption, which typically limit the quality factor (Q factor) [18]-[23]. A possible solution is to merge multiple BICs in the momentum space, which forms a merging BIC. Merging BICs can enhance the Q factor of all nearby resonances in the same band and suppress the out of plane radiation due to their unique topological nature[22]. However, the merging BICs have only been demonstrated in photonics crystal slabs (PCSs) with up-down mirror symmetry, i.e., the slab film is suspended [23]-[25]. On the other hand, many potential applications require up-down asymmetry, including biosensors and microfluidic devices which need a substrate with its refractive index differing from the water environment

[25]-[35]. Therefore, merging BICs without the requirement for up-down symmetry can highly facilitate these applications.

In this paper, we show that merging multiple BICs can be constructed in a metasurface without up-down symmetry. More precisely, we initially illustrate that ultra-high-Q BIC resonances, facilitated by silicon pillars arranged in a Lieb lattice, predominantly exhibit a toroidal dipole mode. By leveraging the topological nature of BICs, we can rearrange the BICs in momentum space and merge them at a specific point. The topological configuration of the merged BICs governs radiative losses of all nearby resonances, making them less susceptible to fabrication imperfections and disorders compared to single BICs [22]. We note that the Lieb lattice is a two-dimensional, edge-depleted square lattice, as illustrated in Figure 1 (BIC region), which is characterized by its flat band structure. Our choice to utilize this lattice stems from its inherent property that simplifies the formation of accidental BICs due to destructive interference. For the fabrication, our BIC metasurface is composed of silicon pillars arranged in a Lieb lattice patterned on a glass substrate. In addition, most BIC metasurfaces require a relatively large footprint to suppress in-plane radiation. However, several applications such as nanolasers [37], optofluidics, and pixelated biosensing [16] applications prefer a small footprint design for better integration and sensitivity. To miniaturize the size of BIC metasurfaces, we enclose the BIC with a photonic band-gap mirror, which can prevent transverse leakage and effectively trap light in-plane [23] [24]. The organization of this paper is as follows. First, we describe the design of the metasurface leveraging the Lieb lattice. We show that a dominant toroidal dipole mode is generated and accidental BICs still stably exist without the up-down mirror symmetry. Then, by tuning the structural parameters, we are able to merge the BICs in the vicinity of the Γ point and get a much better Q factor scaling rule in momentum space. Finally, we introduce a lateral band gap mirror that can significantly improve the Q factor of the BIC metasurface by forbidding in-plane radiation.

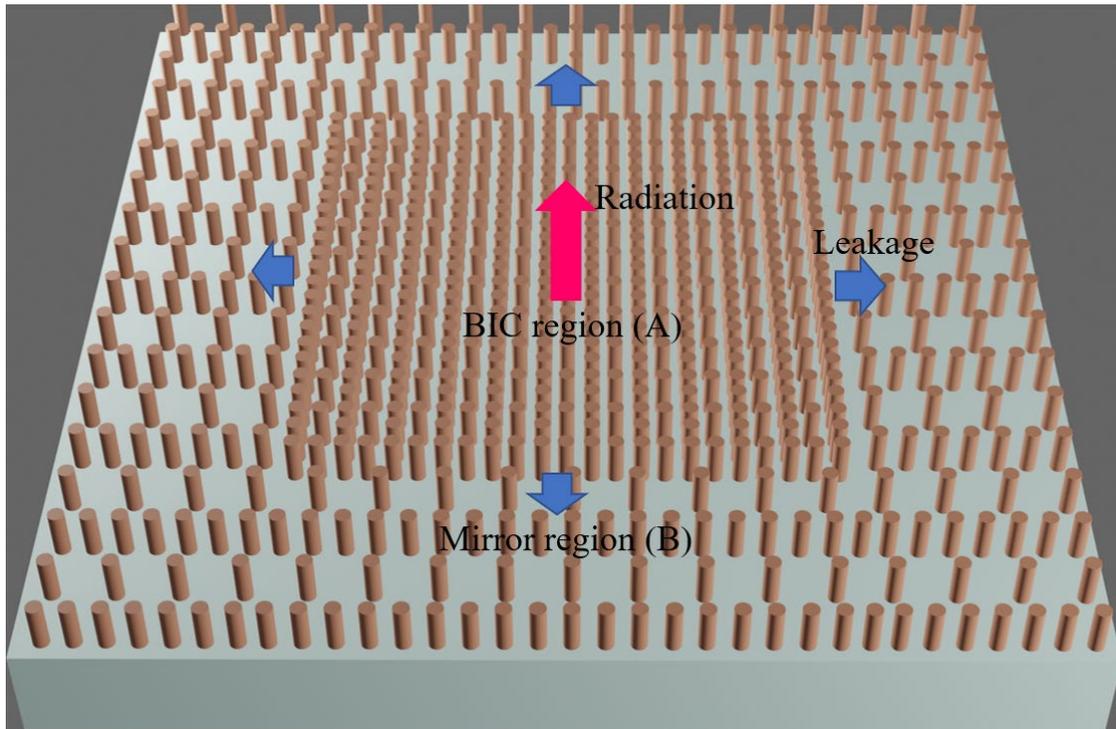

Figure 1. Schematic of a photonic crystal slab and the factors contributing to loss. The metasurface is composed of two regions. Region A: BIC region supporting a BIC resonance. Region B: mirror region for suppressing in-plane radiation.

**Merging Toroidal dipole BICs in Lieb Lattice**

Leveraging the commercial finite element simulation software (COMSOL Multiphysics), we investigate the merging of multiple BICs in a Lieb lattice metasurface. We primarily focus on the TM2 band (illustrated by the blue line in Figure 3a). We choose this particular band due to the existence of eight accidental BICs, which encircle a symmetry-protected BIC (Figure 3b, first panel). The BIC metasurface consists of silicon pillars (n = 3.6) arranged in a Lieb lattice on a glass substrate (n = 1.46). The refractive index for the upper medium (assumed as water) is 1.33. The structural parameters are shown in Fig. 2a. The periodicity is set as 598 nm and the radius is at 90 nm, while the pillar height was tuned to 493 nm to approach the merging BIC condition. Upon visualizing the distribution of the electrical (E) and magnetic (H) fields (refer to Figure 2a), the manifestation of a toroidal dipole mode becomes evident. This mode is defined by a circular H field and a corresponding E field that shapes into a toroidal configuration. We also performed multipole decomposition analysis as described in the SI. The multipole decomposition analysis further validates that the dominant mode is characterized by a toroidal dipole (as shown in Figure 2b).

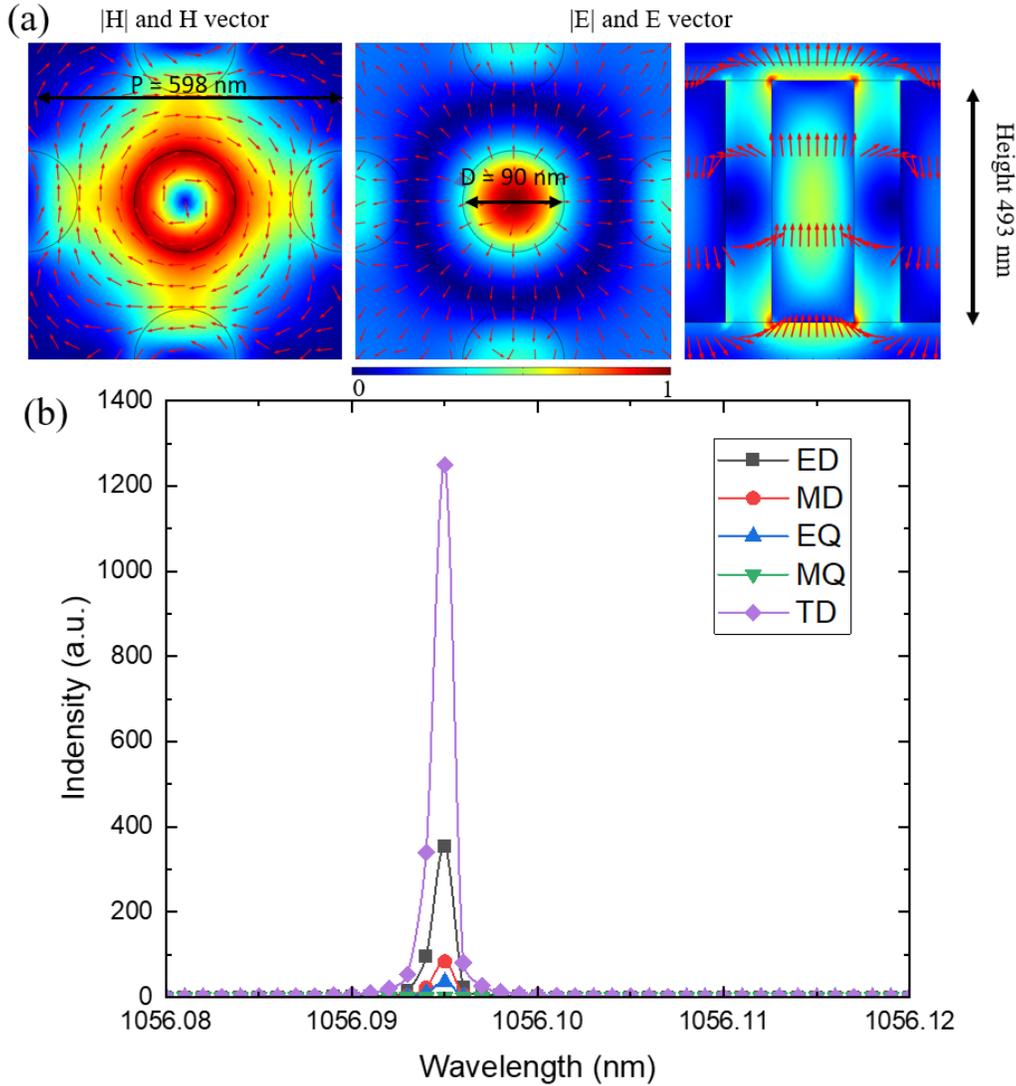

Figure 2. Mode profile and multipole decomposition analysis. The left figure of (a) illustrates the mode profile and vectors of magnetic field in xy cut plane, while the two right figures show the mode profile and vectors of electric field in xy plane and xz cut planes, respectively. Hotspots exposed on the top surface of the pillar show that in this BIC merging metasurface, electrical field is accessible for applications demanding light-matter interactions such as optical trapping and biosensing. (b) Multipole decomposition result indicate that the mode is dominated by a toroidal dipole mode.

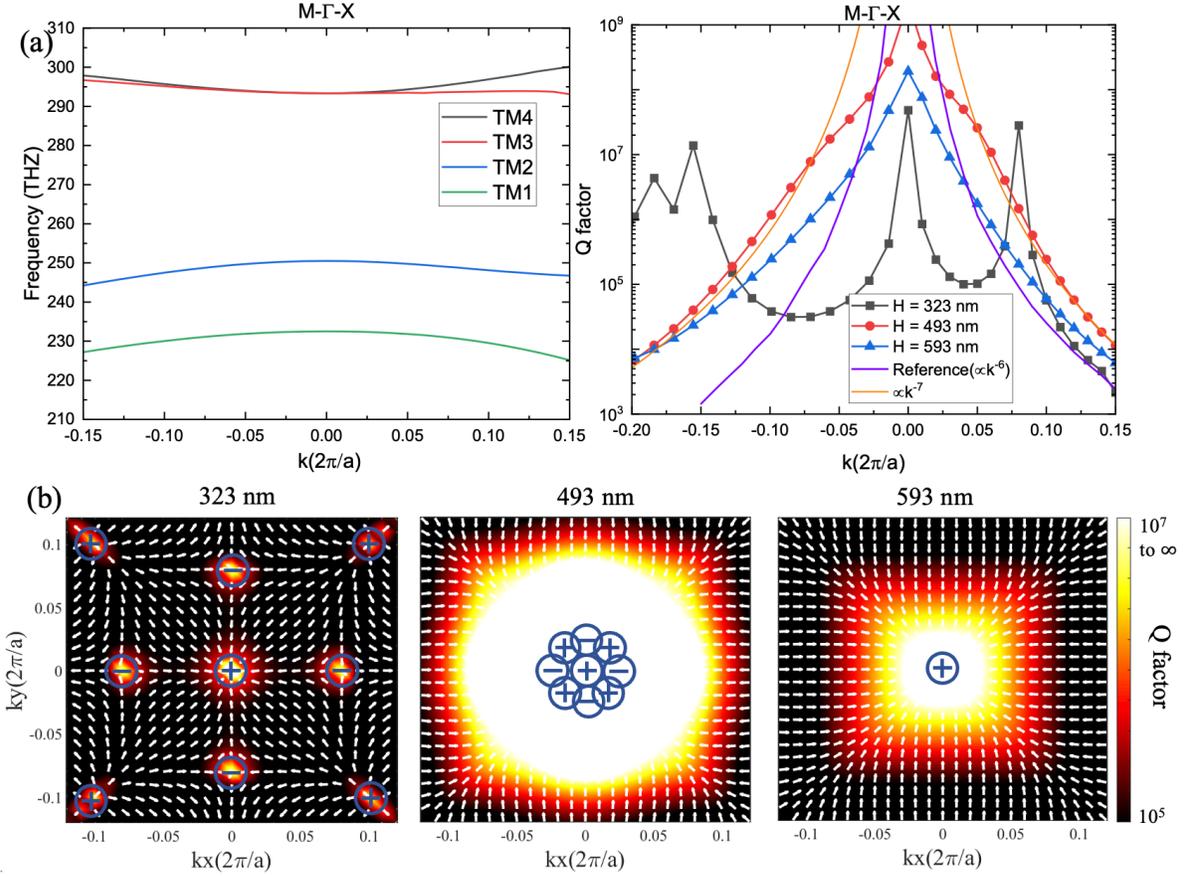

Figure 3. Merging nine BICs in momentum space. (a) Left: Simulated band structure. The TM2 band is marked as blue. Right: Q factors with respect to in-plane k vectors presenting evolution from "before merging" (323 nm;black) to "after merging " (593 nm;blue). The transition (493 nm;red ) is at "in merging" state, which shows higher Q factor than other states along both the $\Gamma -$ X and $\Gamma - M$ directions. This is caused by a change to a scaling rule of $Q$ close to $1/k^7$. (b) Multiple BICs appear on band TM2. The colormap shows simulated Q for various values of the sample height. The arrow shows far-field polarization vector. When height is tuned from 323 nm (left) to 593 nm (right), nine BICs merge into a single BIC. The transition (height = 493 nm) corresponds to the merging-BIC configuration which shows relatively large Q factor.

To understand the topological nature of the proposed BIC system, we calculate the polarization vectors of the far field radiation and the Q factor distributions in momentum space, as depicted in Figure 3b. For the modes above the light cone and wavelengths below the diffraction limit, the nonzero components of the propagating-wave are proportional to the zero-order Fourier coefficient of Bloch wave function. Thus, the far-field polarization vector is defined as the in-plane projection of $\langle \vec{u}_k \rangle$: [2]

$$\vec{c}(\vec{k}) = \hat{x} \cdot \langle \vec{u}_k \rangle + \hat{y} \cdot \langle \vec{u}_k \rangle \quad (1)$$

,where $\langle \vec{u}_k \rangle$ is the zero order Fourier coefficient of Bloch wave function and $\vec{c}(\vec{k})$ is the far-field polarization vector. Since BICs are the zero emission into the far field, they can be defined as the singularity of far-field polarization vector. This fact has been verified both theoretically and experimentally [34] [35]. Singularities (vortices) are characterized by the topological charges, which is the winding number of the polarization vector:[2]

$$q = \frac{1}{2\pi} \oint_C d\vec{k} \cdot \nabla_k \varphi(\vec{k}) \quad (2)$$

,where C is a closed simple path in k space that surrounds the BIC in a counterclockwise direction. $\varphi(\vec{k}) = \arg[c_x(\vec{k}) + c_y(\vec{k})]$ is the angle of the polarization vector of which the x and y components are denoted by $c_x(\vec{k})$ and $c_y(\vec{k})$. The left figure of Figure 3b shows that there is a symmetry-protected BIC with a topological charge +1 at the Γ point and eight accidental BICs with topological charge ±1 in Γ − X and Γ − M directions. The symmetry-protected BIC is fixed at Γ points, while the accidental BICs can move towards the Γ points when we increase the height of the silicon pillar. At height = 493 nm, the eight accidental BIC merges with the symmetry-protected BIC (the middle figure of Figure 3b). If we further increase the height, the BICs with opposite topological charges will annihilate, thus a single BIC with the topological charge +1 persist at height = 593 nm. We note here that the way we merge BICs is to tune the height instead of tuning the periods reported in previous works [23]. This is because the movement of accidental BIC in momentum space would be much slower when we change the height rather than the periods, we can therefore achieve more precise control.

The Q factor distributions in Γ − X and Γ − M directions are shown in Figure 3a. In contrast to the single BICs with topological charge ±1 (height = 323 nm), the merging BIC, attained through the reconfiguration of all nine BICs (height = 493 nm), exhibits significantly higher Q factors across a wide momentum range due to its fundamentally distinct scaling properties. For the single BICs in Γ − X and Γ − M directions, the scaling rule of Q factor roughly obeys $Q \propto 1/(k^2(k^2 - k_{BIC}^2)^2)$, where $k$ is the in-plane wavevector. For the merging BIC, the scaling rule we obtained is of approximately $Q \sim 1/k^7$ ( as shown by the yellow curve in the right panel of Figure 3a). Compared with the scaling rule $Q \sim 1/k^6$ reported in previous works [23], our design based on the Lieb lattice metasurface provides a superior scaling performance. As delineated in Figure 3b, we have successfully demonstrated that our merged BIC can be achieved without necessitating up-down mirror symmetry. This discovery is of significant consequence for metasurface biosensors and optofluidic applications, given that the refractive index of the superstrate medium frequently differs from the substrate in these specific contexts.

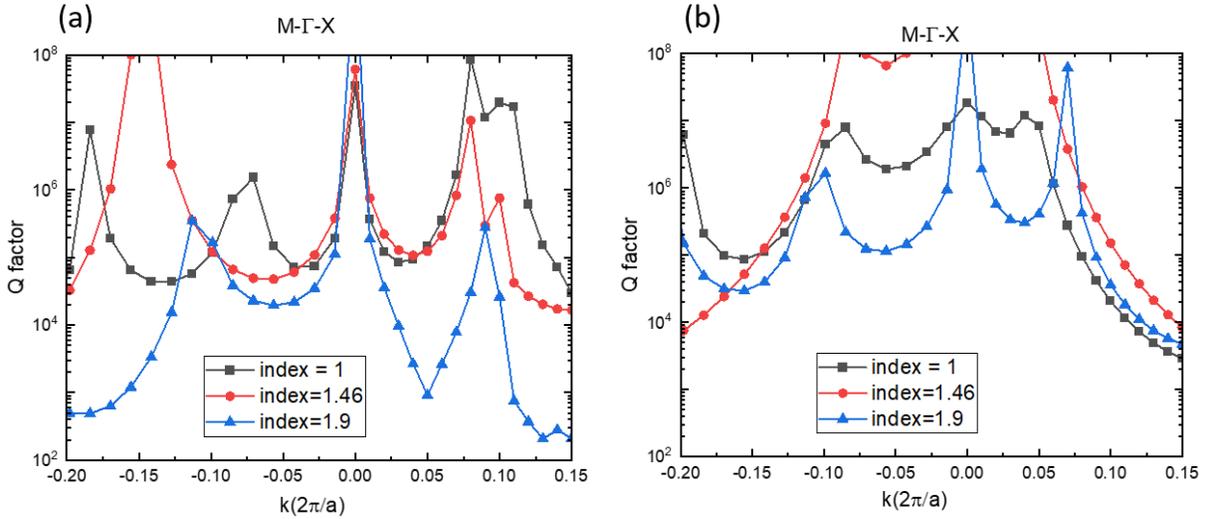

Figure 4. Merging BIC when varying the refractive index of the top media. (a) shows that at different refractive index, there are still nine BICs in the vicinity of Γ point, but they are at different positions in momentum space. (b) shows that if we tune the height at 530 nm, a merging BIC forms when refractive index = 1.46. And the accidental BICs also move towards center for refractive index = 1 and 1.9. he. Their merging can still be achieved by further tuning the height of pillar. (a) and (b) indicate that the merging BIC is robust to refractive index changes varies.

**Merging BIC with tunable refractive index of top layer**

The aforementioned results in the preceding section show that merging BIC can be formed when the refractive index of the top layer is 1.33. Here, we show that for a wide range of refractive index of the top layer between 1 to 1.9, we can still achieve the merging BIC by tuning the height of the pillar in the Lieb lattice. Figure 4a shows that accidental BICs are preserved in the $\Gamma - X$ and $\Gamma - M$ directions when the refractive of the top layer varies in the range of 1 to 1.9, while the refractive index of substrate is kept at 1.46 and the height is set as 323 nm. By increasing the height of the pillar, the eight accidental BICs can move towards the symmetry-protected BIC at the center (just as the evolution shown in Figure 4b). When the refractive index of top layer is set to 1.46, the merging BIC is approached at a height of 530 nm. Moreover, the accidental BICs for refractive index 1 and 1.9 both move towards the center. These results indicate our merging BIC is robust to the refractive index mismatch between the superstrate and the substrate. This is because no matter what the refractive index of the top layer is, we can reconfigure the eight BICs towards the vicinity of symmetry-protected BIC forming a merging BIC. In simpler terms, merging BICs by this design can be attained with a wide range of refractive indices for the top layer.

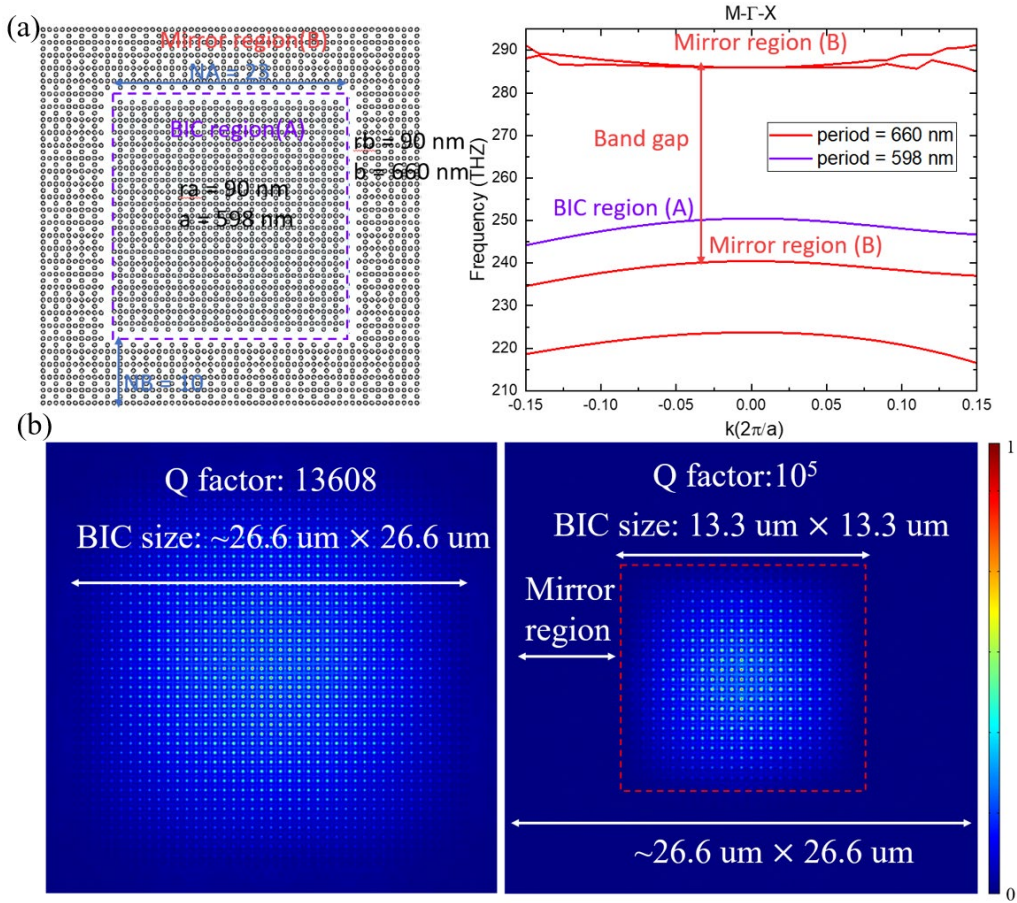

**Figure 5.** Schematic of the band gap mirror and finite size structure simulation. BIC region A of Na × Na array of Lieb lattice pillars (Na = 23, a = 598 nm, ra = 90 nm) is encircled by boundary region B of Nb × Nb array ( Nb = 10, b = 660 nm, rb = 90 nm). The right figure on the top shows the band structures of region A and B in $\Gamma - X$ and $\Gamma - M$ directions (simulated by COMSOL Multiphysics). TM2 band in region A (purple) is embedded in the band gap of region B (red). The left figure on the bottom line shows the Q factor and field distribution of a finite size structure without lateral mirrors (footprint 26.6×26.6 um), and the right figure shows the Q factor and field distribution of a finite size structure surrounded by a 10×10 band gap mirror. The one with a mirror boosts the Q factor up to $10^5$.

**Miniaturizing the array of Lieb lattice BIC metasurface**

In a finite-sized BIC metasurface, the Q factor is limited by the in-plane scattering loss. To address this limitation, we introduce a band gap mirror that encircles the BIC structure, effectively suppressing radiation by prohibiting wave propagation within the mirror. This allows the footprint of the BIC metasurface to be scaled down while maintaining the same Q factor as that of a larger array size. The left figure in Figure 5a illustrates the design, featuring a mirror region outside the BIC region arranged in a Lieb lattice with silicon pillars matching the height of the BIC region. This design minimizes the fabrication difficulties. The right figure displays the band structure of the mirror. By altering the period of the mirror region with respect to the BIC region, we ensure that the band of the BIC mode falls within the band gap of the mirror region. Consequently, in-plane propagating waves are unable to propagate laterally and are perfectly reflected. Numerical simulation results conducted using COMSOL Multiphysics are presented in Figure 5b. The BIC-mirror combined structure retains the same footprint as the independent BIC structure yet exhibits an approximately 10-fold increase in the Q factor. This approach enables the design of BIC structures with a small footprint, which holds great promise for nanolasers, biosensors, and quantum computing.

**Conclusion**

In conclusion, we have proposed a novel mechanism for merging BICs at the Γ point, even in the absence of up-down symmetry. To accomplish this, we utilize a toroidal dipole BIC implemented in a Lieb lattice metasurface. These accidental BICs, located away from the Γ point can stably exist within a wide range of environment refractive indices. By incorporating a lateral band gap mirror made of a Lieb lattice, the size of the BIC metasurface can be effectively reduced without compromising the Q factor. The Lieb lattice band gap mirror prevents in-plane radiation, ensuring the preservation of high Q factors. Our discovery offers substantial potential for integrating BICs into various applications, including nanolasers, optical trapping, biosensors, and quantum computing. This is particularly pertinent in scenarios where high-Q miniaturized devices are sought. Furthermore, due to the inherent topological nature of BICs, our approach provides robustness, paving the way for further enhancing the performance of optoelectronic devices.

**AUTHOR INFORMATION**
Corresponding Author
Justus C. Ndukaife − Vanderbilt Institute of Nanoscale Science and Engineering and Department of Electrical Engineering and Computer Science, Vanderbilt University, Nashville, Tennessee 37235, United States; orcid.org/0000-0002-8524-0657; Email: justus.ndukaife@vanderbilt.edu
ORCID
Guodong Zhu: 0009-0004-1659-3522
Sen Yang: 0000-0002-0056-3052
Justus C. Ndukaife: 0000-0002-8524-0657

**Notes**
The authors declare no competing financial interest.

**ACKNOWLEDGMENTS**
The authors acknowledge financial support from the National Science Foundation (NSF ECCS-1933109) and Vanderbilt University.

**For Table of Contents Use Only**

**Merging toroidal dipole bound states in the continuum without up-down symmetry in Lieb lattice metasurfaces**
Guodong Zhu, Sen Yang and Justus C. Ndukaife

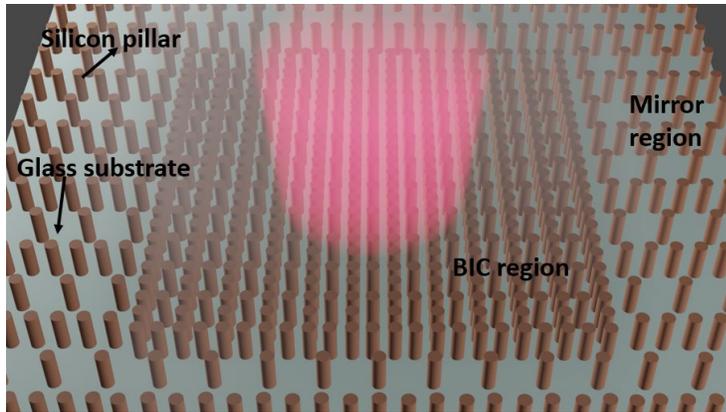
Table of Contents (TOC) Graphic: this schematic diagram shows compact BIC system described in this manuscript. It consists of two parts: BIC region and Mirror region. The only deference of these two regions is the periodicity. The yellow represents the silicon pillar, while the white represents the glass substrate. The red indicates the radiation.